\documentclass[
aps,
pra,
showpacs
amsmath,
amssymb,
onecolumn,
floatfix,
longbibliography,
letterpaper,
lengthcheck,
superscriptaddress
]{revtex4-2}

\usepackage{graphicx}
\usepackage{floatflt,epsfig} 
\usepackage{mathtools}
\usepackage{color}
\usepackage{braket}
\usepackage{hyperref}
\usepackage[normalem]{ulem}
\usepackage[english]{babel}
\usepackage{blindtext}
\usepackage{lipsum}  
\usepackage{amsthm}
\usepackage{tabularx}
\usepackage{bm}
\usepackage{dsfont}
\usepackage{bbold}
\usepackage{ulem}
\usepackage{mwe,tikz}
\usepackage[percent]{overpic}
\usepackage{todonotes}

\begin{document}

\title{Dynamic of Single Molecules in Collective Light-Matter States from First Principles}

  \author{Christian Sch\"afer}
  \email[Electronic address:\;]{christian.schaefer.physics@gmail.com}
  \affiliation{Department of Microtechnology and Nanoscience, MC2, Chalmers University of Technology, 412 96 G\"oteborg, Sweden}

\date{\today}

\begin{abstract}
The coherent interaction of a large collection of molecules with a common photonic mode results in strong light-matter coupling, a feature that proved highly beneficial for chemistry and termed the research topics polaritonic and QED chemistry.
Considering complex microscopic chemical reactions in combination with a macroscopic number of molecules renders existing \textit{ab initio} approaches inapplicable.
In this work, I introduce a simple approach to capture the collective nature while retaining the full \textit{ab initio} representation of single molecules. By embedding the majority of the molecular ensemble into the dyadic Green tensor, we obtain a computationally cheap and intuitive description of the dynamic of a single molecule in the ensemble --- an approach that seems ideal for polaritonic chemistry. 
The introduced embedding radiation-reaction potential is thoroughly discussed, including prospects, applications and limitations. A first application demonstrates the linear response of single molecules that are part of a larger ensembles of molecules. Then, by virtue of a simple proton-tunneling model, I illustrate that the influence of collective strong coupling on chemical reactions features a nontrivial dependence on the number of emitters. Bridging classical electrodynamics, quantum optical descriptions and the \textit{ab initio} description of realistic molecules, this work can serve as guiding light for future developments and investigations in the quickly growing fields of QED chemistry and QED material design.
\end{abstract}

\date{\today}

\maketitle


\section{Introduction}

The self-consistent interaction between light and matter developed over the recent years into a viable and important tool to non-intrusively shape chemistry and materials on demand. 
Specific resonator geometries (cavities) support a limited set of eigenmodes which couple strongly to matter resulting in new quasiparticles, so called polaritons. Typical realizations of those resonators range between nanometer sized plasmonic structures, that can couple to individual molecules, and Fabry-P\'erot cavities in which many molecules couple collectively \cite{ebbesen2016,garcia2021manipulating,simpkins2021mode,sidler2021perspective,forn2019ultrastrong,kockum2019ultrastrong}.
Such a hybridization between cavity mode(s) and matter exists also in the absence of driving, in contrast to Floquet physics which is often limited by induced heating and decoherence \cite{schafer2018insights,sato2020floquet} --- the mere existence of the confined modes can alter materials.
This includes the control of photo-chemical reactions by coupling to electronic excitations \cite{hutchison2012,chikkaraddy2016,galego2016,munkhbat2018suppression,fregoni2018manipulating,groenhof2019tracking,kowalewski2016cavity,Vendrell2018,fabri2021probing,li2022semiclassical}, 
ground-state chemical reactions via vibrational strong coupling \cite{thomas2016,thomas2019tilting,Lather2019,schafer2021shining,li2021collective,li2021cavity}, 
and energy- and charge-transfer over macroscopic dimensions \cite{coles2014b,feist2015,orgiu2015,zhong2017energy,schafer2019modification,du2018theory,hagenmuller2018}.
Furthermore, notable progress in the construction of cavities \cite{wrigge2008efficient,bajoni2008polariton,chikkaraddy2016,wang2017coherent,baranov2019ultrastrong,hubener2021engineering} has lead to a steadily rising number of applications outside chemistry, ranging from polariton-mediated lasing \cite{deng2003polariton,christopoulos2007room,kena2010room} over material design \cite{slootsky2014room,graf2017electrical,latini2021ferroelectric,schlawin2022cavity} to quantum information theory \cite{ghosh2020quantum,luders2021quantifying,wasielewski2020exploiting}.

However, most of the so far existing chemically relevant applications couple many molecules collectively to a common cavity mode.
On one hand, a predictive theoretical description of chemical processes requires a thorough description of the electronic and vibrational structure from first principles. On the other hand, a direct description of large ensembles is prohibited by the quickly increasing computational cost. 
The vast majority of theoretical descriptions are therefore based on simplified models that posses a limited applicability to complex systems and chemical reactions.
Descriptions from first principles have been restricted to small numbers of molecules \cite{schafer2021shortcut,tancogne2019octopus,noda2019salmon,mennucci2019multiscale,luk2017multiscale,fregoni2021strong,bonini2021ab}. Notable representatives are quantum-electrodynamical density-functional theory (QEDFT) \cite{tokatly2013,ruggenthaler2014,schafer2021making,PhysRevLett.121.113002} and cavity coupled-cluster theory \cite{haugland2020coupled,haugland2020intermolecular}.

By embedding the majority of the molecular ensemble into a local potential, we will discover a path to open first-principles techniques towards collective strong-coupling with arbitrary numbers of ensemble molecules and species.
This approach represents an extension of the radiation-reaction ansatz derived in reference \cite{schafer2021shortcut} and inherits its computational and conceptual simplicity, allowing an almost effortless implementation into existing time-dependent density-functional theory (TDDFT) libraries.

We start by introducing the radiation-reaction ansatz (section~\ref{sec:embedding}) that accounts for the fully self-consistent interaction between classical light and quantum matter in a simple and elegant fashion. Based on this construction and the experimental conditions in polaritonic chemistry, the majority of the ensemble will be embedded into the local potential (section~\ref{sec:collective}). Solving the common Kohn-Sham equations \cite{hohenberg1964,kohn1965,gross1984} with the addition of this embedding radiation-reaction potential provides then access to the dynamic of single molecules in collective light-matter states. Its application and physical consequences are demonstrated for electronic and vibrational strong coupling (section~\ref{sec:application}).

\section{Embedding the ensemble into the electromagnetic environment}\label{sec:embedding}
In the non-relativistic limit, the Hamiltonian governing the electronic and nuclei dynamic under the influence of classical electromagnetic fields is given by
$ \hat{H} = \sum_i \frac{1}{2m_i}\left( -i\hbar\nabla_i - q_i \textbf{A}(\textbf{r}_i t)/c \right)^2 + \hat{H}_\parallel + \varepsilon_0/2\int d\textbf{r}[ \textbf{E}_{\perp}(\textbf{r}t)^2 + c^2\textbf{B}(\textbf{r}t)^2] $ with fixed Coulomb gauge $\nabla \cdot \textbf{A}=0$. In addition, the self-consistent interaction between light and matter dictates that the electromagnetic fields follow Maxwell's equation.
As demonstrated in reference \cite{schafer2021shortcut}, the classical interaction with light can be efficiently incorporated into electronic structure theory via the transverse component of the dyadic Green tensor $\textbf{G}_\perp$ 
\begin{align*}
\textbf{E}_{r,\perp}(\textbf{r},\omega) = i\mu_0\omega\int_V dr' \textbf{G}_\perp(\textbf{r},\textbf{r}',\omega) \cdot (-e\textbf{j}(\textbf{r}',\omega))
\end{align*}
which is the formal solution of Helmholtz's equation
\begin{align}\label{eq:helmholtz}
\big[\nabla \times \frac{1}{\mu_r(\textbf{r}\omega)}\nabla \times - \omega^2 \mu_0\varepsilon_0\boldsymbol\varepsilon_r(\textbf{r}\omega)\big] \textbf{G}(\textbf{r},\textbf{r}',\omega) = \boldsymbol\delta(\textbf{r},\textbf{r}')
\end{align} 
and characterizes the electromagnetic environment. The microscopic paramagnetic current $\textbf{j}$ serves as driving inhomogeneity, i.e., oscillating charges emit light. In addition, we have the freedom to assume that parts of the system behave as local and potentially isotropic linear medium $\boldsymbol\epsilon_r,~\mu_r$ which shape via the Helmholtz equation the electromagnetic environment, represented by $\textbf{G}$. 
\footnote{The relative permeability for the systems of interest in this work is negligible and we will assume $\mu_r \approx 1$.} 
In this way, $\textbf{G}$ can subsume large fractions of a system while we free computational resources for the microscopic system described by $\textbf{j}$. The local radiation-reaction potential 
\begin{align}
\begin{split}
\label{eq:rrpotential}
\hat{V}_{rr}(t)&=-\hat{\textbf{R}}\cdot\textbf{E}_{r,\perp}(t)\\
\textbf{E}_{r,\perp}(t) &= \big[\mathcal{F}_t^{-1}(i\mu_0 \omega \textbf{G}_\perp (\omega)) \ast \int dr (-e \textbf{j}(\textbf{r}t))\big]
\end{split}
\end{align}
acts on the microscopic currents via the Schr\"odinger equation and manifests the self-consistent interaction with transverse fields within the long-wavelength approximation. This allows us to extend for example TDDFT with the self-consistent classical interaction between an electromagnetic environment and matter by simply adding $v_{rr}(\textbf{r}t)=e\textbf{r}\cdot \textbf{E}_{r,\perp}(t)$ to the usual Kohn-Sham equations. Since we anyway have access to the dipolar moment in TDDFT and $\int dr (-e\textbf{j}(\textbf{r}t)) = \partial_t \textbf{R}(t)$, the radiation-reaction potential is trivial to implement into existing TDDFT libraries.

The longitudinal components of the electric field follow in dipolar approximation the equivalent $ \hat{V}_{\parallel}(t)=-\hat{\textbf{R}}\cdot\textbf{E}_{r,\parallel}(t),~\textbf{E}_{r,\parallel}(\textbf{r},\omega) = i\mu_0\omega\int_V dr' \textbf{G}_\parallel(\textbf{r},\textbf{r}',\omega) \cdot (-e\textbf{j}(\textbf{r}',\omega))$ system of equations. The longitudinal components are most relevant only in the near-field, i.e., they can be safely ignored when considering the cavity-mediated action between emitters that are separated by many nanometers. Our focus will rest in the following on the collective strong coupling mediated by transversal cavity fields. Nevertheless, solvation and other near-field effects such as the interaction with plasmonic systems are often critical in chemistry \cite{mennucci2012polarizable,chikkaraddy2016,munkhbat2018suppression,kumar2019plasmon} and a wholesome picture should consider longitudinal and transversal fields on equal footing.
We will focus in the following on electromagnetic environments that embody Fabry-P\'erot-like eigenmodes and a large number of emitters. Clearly, $\textbf{G}$, and with it the presented embedding radiation-reaction ansatz, is generic and able to account for any kind of complex environment \cite{schafer2021shortcut}.

In the collective strong-coupling regime between many molecules and a cavity, single photon processes dominate and the individual effect of the cavity field acting on a single molecule is comparably weak. The electronic and nuclear excitation structure remains then largely unaffected, the combined light-matter system hybridizes based on those bare excitations. As long as this condition is satisfied it is safe to separate the matter-ensemble into two parts. The first describes a single explicit molecule featuring the paramagnetic current $\textbf{j}$, which will be obtained as solution to Schr\"odinger's equation. The second accounts for all the remaining molecules as polarizable material $\boldsymbol\varepsilon_r(\textbf{r},\omega)=\textbf{1}+\boldsymbol\chi_E(\textbf{r},\omega)$. Surely this separation is also convenient to account for two different species, e.g., the reactive molecule and its solvent. 
In this sense, the problem of chemical reactions in collective light-matter coupling can be seen as impurity problem, a single molecule undergoing a reaction is influenced by a large electromagnetic environment.

Knowing the bare Green tensor $\textbf{G}_0$ (empty cavity), we can obtain the 'ensemble-dressed' dyadic Green tensor with the help of the Dyson equation
\begin{align}
\begin{split}
\label{eq:dyson}
&\textbf{G}(\textbf{r},\textbf{r}',\omega) = \textbf{G}_0(\textbf{r},\textbf{r}',\omega)\\ 
&+ \frac{\omega^2}{c^2} \int_{V}dr'' \textbf{G}_0(\textbf{r},\textbf{r}'',\omega) \cdot \boldsymbol\chi_E(\textbf{r}'',\omega) \cdot \textbf{G}(\textbf{r}'',\textbf{r}',\omega)~.
\end{split}
\end{align}
We assumed here (and earlier) that the generic susceptibility $ \textbf{P}(\textbf{r},\omega) = \varepsilon_0 \int dr^3 \boldsymbol{\chi}(\textbf{r},\textbf{r}',\omega) \cdot \textbf{E}(\textbf{r}',\omega) $ is local $\boldsymbol{\chi}(\textbf{r},\textbf{r}',\omega) \approx \boldsymbol\chi(\textbf{r},\omega) \delta(\textbf{r}-\textbf{r}')$. In many situations, the orientation of emitters within a gas or fluid will be random and its average polarizability is consequentially isotropic. 
\footnote{For non-local response, e.g., potentially next to plasmonic surfaces or within aggregates, the above equation can be generalized using $\boldsymbol\chi(\textbf{r},\textbf{r}',\omega)$. Its ultimate usefulness depends however on the description of $\boldsymbol\chi_E$ and it remains challenging to this date to capture complex non-local response from first principles.}
The concentration of the ensemble of emitters can vary such that the above equation would require for instance an iterative solution. We can, however, obtain an analytic solution in the special case that the ensemble occupies a small region in space compared to the wavelength's supported by $\textbf{G}_0$. 
\footnote{Such an iterative procedure would start from $\textbf{G}\approx \textbf{G}_0$. The convolution is simplified in Fabry-P\'erot cavities by the orthogonal normalmode structure $\int dx'' \sin(\pi x'' n_x/L)\sin(\pi x'' n_x'/L)\int\int dy''dz''=V/2 \delta_{n_x n_x'}$ and $ \int dy dz = V/L $. The main text illustrates a simplified motivation but we can analytically obtain the same equation for a single Fabry-P\'erot mode by either truncating the Dyson equation at first order or assuming that the spatial structure of $\textbf{G}$ remains proportional to $ \sin(\pi x'' /L)\sin(\pi x' /L) $. The integration and subsequent inversion of the equation leads then to the same expression if we assume that the $N_E$ ensemble molecules are concentrated to half the cavity volume. The long-wavelength approximation leads therefore to qualitatively consistent results but the precise choice of how $\chi_E$ is distributed in space results in slightly modified coefficients.} 
The following sections demonstrate the simplicity of the embedding radiation-reaction ansatz and illustrate its physical consequences for the collective strong-coupling of electronic and vibrational excitations.

\subsection{The Radiation-Reaction Potential for Collective Strong Coupling}\label{sec:collective}

For a dilute ensemble satisfying the long wavelength approximation, the response can be approximated as
$\boldsymbol\chi_E(\textbf{r}'',\omega) \approx \boldsymbol\chi_E(\textbf{r}_0,\omega)\theta(\textbf{r} \in V_{E}) \approx \boldsymbol\chi_E(\omega)  V_{E} \delta(\textbf{r}-\textbf{r}_0)$. Then, 
\begin{align}
\label{eq:gdyson2}
\textbf{G}(\omega) \approx \textbf{G}_0(\omega)+ V_E \frac{\omega^2}{c^2} \textbf{G}_0(\omega)\cdot \boldsymbol\chi_E(\omega) \cdot \textbf{G}(\omega)
\end{align}
is evaluated at the position of the explicit molecule and $V_E$ is the volume occupied by the ensemble of emitters.
The remaining task is to identify $\boldsymbol\chi_E(\omega)$, in the following exemplified for TDDFT.

Density-functional theory is designed to handle longitudinal perturbations, transversal perturbations imply the description of transverse currents which in turn implies the usage of current-density functional-theory \cite{vignale1987density,vignale1996current,de2001current,maitra2003current}. In the long-wavelength approximation however, Maxwell's equations simplify tremendously and the coupling to longitudinal and transversal fields obtains the same form  $\hat{V}_{\perp,\parallel}(t) = -\hat{\textbf{R}}\cdot \textbf{E}_{\perp,\parallel}(t)$.
The (dipolar) polarization can be conveniently obtained from the density-density response function \cite{ullrich2011} as
\begin{align*}
\textbf{R}(\omega) &= \int dr (-e)\textbf{r} \delta \rho(\textbf{r}\omega) = \boldsymbol{\alpha} \cdot \delta\textbf{E}_{local}(\omega)\\
&= \int dr \int dr' (-e^2)\textbf{r} \chi_{\rho\rho}(\textbf{r},\textbf{r}',\omega) \textbf{r}' \cdot \delta\textbf{E}_{local}(\omega)~.
\end{align*}
We can find a direct relation between polarization density $\textbf{P}(\omega)$ used in Maxwell's equation ($\boldsymbol\chi(\textbf{r},\omega) \approx \boldsymbol\chi(\omega)$) and the dipole $\textbf{R}(\omega)$ calculated from TDDFT
\begin{align}
\begin{split}
\textbf{R}(\omega) &= \boldsymbol\alpha(\omega) \cdot \delta\textbf{E}_{local}(\omega)\\
\textbf{P}(\omega) &= \frac{1}{V_E}\sum_{i=1}^{N_E} \textbf{R}_i(\omega) = \varepsilon_0 \boldsymbol\chi_E(\omega) \cdot \textbf{E}(\omega)\\
\boldsymbol\chi_E(\omega) &= N_E V_E^{-1}  \varepsilon_0^{-1} \boldsymbol\alpha(\omega),~\text{if}~\textbf{E} \approx \delta\textbf{E}_{local}~.
\end{split}
\end{align}
This approximation is reasonable if we focus on weak external fields that are consistent with a linear response treatment of the ensemble response. Furthermore, the long-wavelength approximation implies that the local fields acting on the microscopic current $\textbf{j}$ and the ones acting on the ensemble polarizability $\boldsymbol\alpha(\omega)$ are comparable.
In general, the local fields polarizing a single molecule consist of the sum of the external field plus the polarization-field of the surrounding material. Obtaining the self-consistent relation between micro- and macro-fields is in general nontrivial, posing for instance notable challenges in solid state systems \cite{resta2007theory}. The Clausius-Mossotti relation \cite{kittel1967introduction} provides a simple and reliable solution for isotropic ($\boldsymbol\alpha(\omega) \approx \alpha(\omega) = \frac{1}{3} \text{tr}(\boldsymbol\alpha(\omega))$) dielectric gases and fluids comprising of dilute molecules with negligible permanent dipole
$\chi_E(\omega) = \frac{N_E V_E^{-1} \alpha(\omega) \varepsilon_0^{-1}}{1-\frac{1}{3}N_E V_E^{-1} \alpha(\omega) \varepsilon_0^{-1}}$. Under the assumption of a dilute molecular environment ($N_E/V_E \ll 1$) we obtain again $V_E \chi_E \approx N_E \varepsilon_0^{-1} \alpha(\omega)$. 

Obtaining the polarizability $\boldsymbol\alpha(\omega)$ is a standard task for any time-dependent electronic-structure code. However, it might be not always possible or desirable to obtain the embedding kernel from first principles and instead simplified approaches such as the Drude-Lorentz model present easily accessible alternatives. The following sections will demonstrate the usage of the simple model as well as the full \textit{ab initio} description.

In combination with the radiation-reaction potential $ \hat{V}_{rr}(t)=-\hat{\textbf{R}}\cdot\textbf{E}_{r,\perp}(t) =-\hat{\textbf{R}}\cdot \big[\mathcal{F}_t^{-1}(i\mu_0 \omega \textbf{G}_\perp (\omega)) \ast \int dr (-e \textbf{j}(\textbf{r}t))\big]  $,
the feedback between the ensemble-dressed electromagnetic environment and the local molecule leads to collective states and adjusted local dynamics. The number of ensemble emitters does not influence the computational cost of this approach, the embedding scheme allows us therefore to reach the alleged macroscopic number of molecules participating in experiments.

The simplicity of the resulting equation becomes apparent from a simple example that will follow us throughout this work. For this purpose, we chose a simplified one-dimensional Fabry-P\'erot cavity
$\textbf{G}_{0,\perp} (x,x',\omega) = \sum_{\textbf{k}} \frac{S_{\textbf{k}}(x) S_{\textbf{k}}(x') }{k^2-(\omega/c)^2} \boldsymbol\epsilon_c \boldsymbol\epsilon_c^T$ with the eigenmodes $S_{\textbf{k}}(\textbf{r}) = \sqrt{2/V}\sin(k x),~k=\pi n_x/L_x,~n_x\in \mathbb{N}$ and $\boldsymbol\epsilon_c \perp \textbf{k}$. 
The approximate analytic solution for such a transverse Green dyadic is easily obtained as
\begin{align}
\textbf{G}_\perp^{-1}(\omega) &= \textbf{G}_{0,\perp}^{-1}(\omega) - \frac{\omega^2}{c^2} V_E \boldsymbol \chi_E(\omega) \notag\\
\textbf{G}_\perp(\omega) &\approx g(\omega) \boldsymbol\epsilon_c \boldsymbol\epsilon_c^T\\
\textbf{G}_\perp(\omega) &\approx \boldsymbol\epsilon_c \boldsymbol\epsilon_c^T \left(g^{-1}_0(\omega) - \frac{\omega^2}{c^2} V_E \boldsymbol\epsilon_c^T \cdot \boldsymbol\chi_E(\omega) \cdot \boldsymbol\epsilon_c \right)^{-1}.\notag
\end{align}
We will add a weak cavity-decay rate $\eta$ to account for a finite spectral width (see Appendix for numerical details).
The poles of the Green tensor are then shifted to the positions of the collective ensemble+cavity hybridized states, routinely called polaritonic states, with hybridization energy $\Omega \propto \sqrt{N_E}$ as detailed in the following. 

\begin{figure}[ht]
\includegraphics[width=1.0\columnwidth]{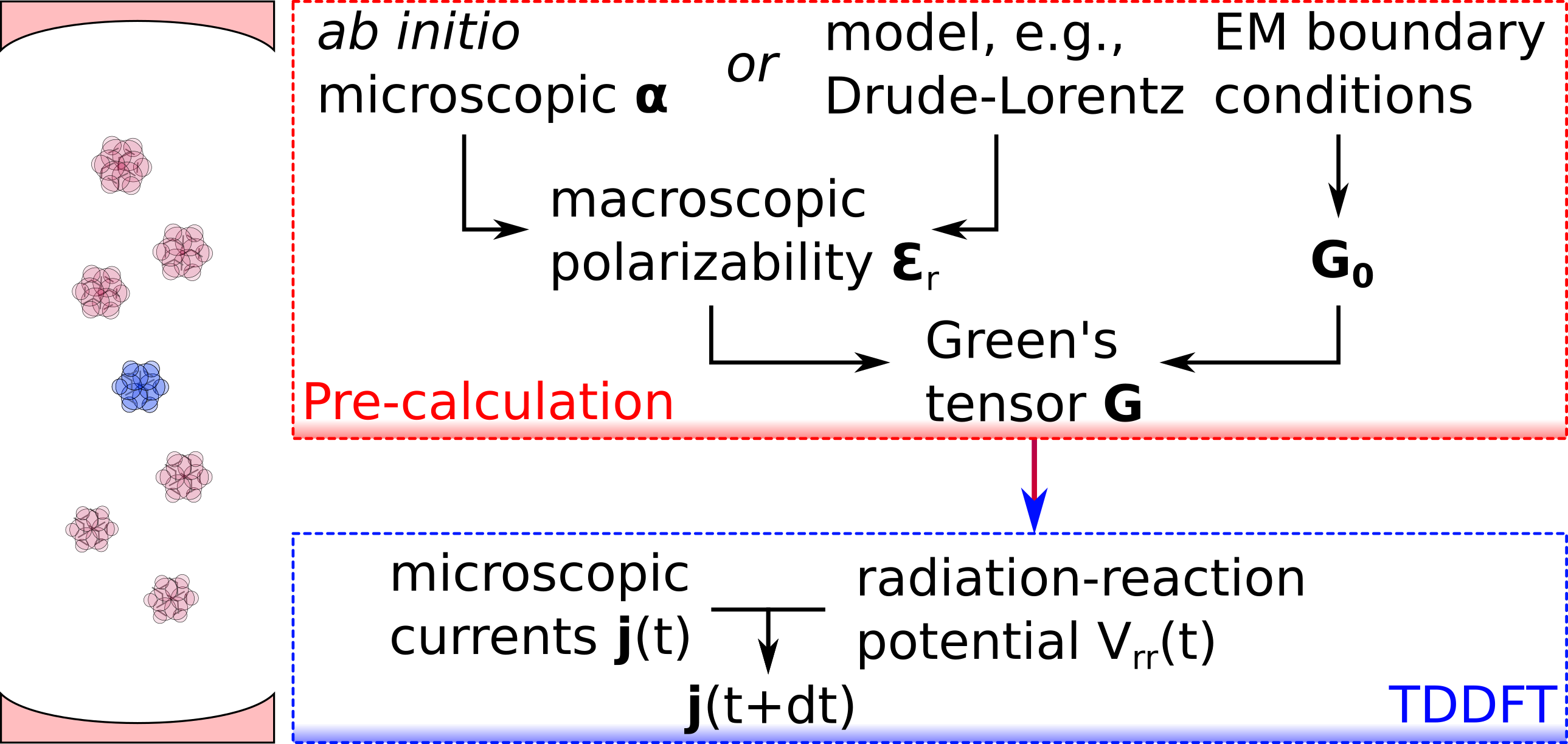}
\caption{Schematic illustration of the embedding radiation-reaction approach exemplified for TDDFT. A single molecule described with time-dependent density-functional theory is coupled to an electromagnetic environment dressed by an ensemble of molecules. This paves the way to describe the dynamic of single molecules in large systems that are collectively coupled to electromagnetic environments from first principles.}
\label{fig:concept}
\end{figure}

The necessary computational steps for a real-time calculation of the molecular dynamics in collective light-matter states is illustrated in figure~\ref{fig:concept} and comprises:
(1) Obtain ground-state density (potentially including QEDFT corrections).
(2) Calculate or model $\boldsymbol\chi_E(\omega)$ (e.g. via $\boldsymbol\alpha(\omega)$).
(3) Obtain dyadic Green tensor, either by solving for $\textbf{G}_0$ and dressing it according to equation \eqref{eq:dyson}/\eqref{eq:gdyson2} or directly solving equation~\eqref{eq:helmholtz} with $\boldsymbol\varepsilon_r(\textbf{r},\omega)$.
(4) Perform real-time TDDFT propagation as usual including the embedding radiation-reaction potential equation~\eqref{eq:rrpotential} and potential quantum-corrections from QEDFT as exemplified in App.~\ref{app:lda}. The radiation-reaction potential is trivially combined with existing TDDFT libraries as the integrated current is given via the continuity equation as $\int dr (-e\textbf{j}(\textbf{r}t)) = \partial_t \textbf{R}(t)$ and therefore naturally obtained during the propagation. 

The additional steps 2-3 are performed only once before any real-time propagation, their computational cost remains therefore small and is dominated by obtaining $\boldsymbol\alpha(\omega)$.

Equipped with the necessary framework, we are now able to describe the dynamic of single molecules in collective strong-coupling. We start by demonstrating the emergence of collective states for small artificial but also realistic systems. We then shift our focus to chemistry and illustrate for a simple proton-tunneling model how chemical reactions can be described and modified via vibrational strong-coupling.

\section{Application of the embedding radiation-reaction potential}\label{sec:application}

\subsection{Illustration of the Single-Molecule Response}

\begin{figure}[ht]
\includegraphics[width=1.0\columnwidth]{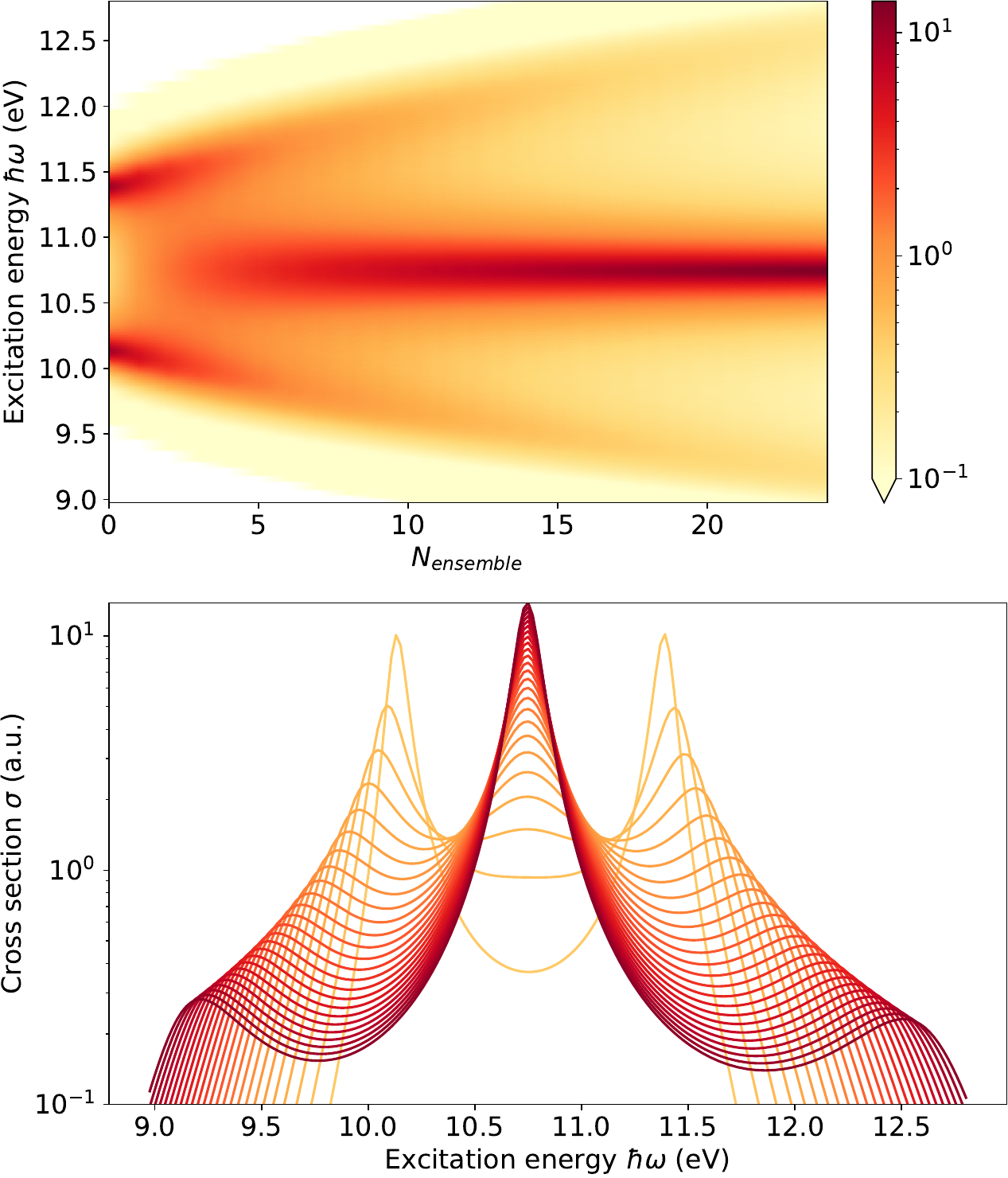}
\caption{Photoabsorption cross-section $\sigma(\omega)$ of one-dimensional hydrogen coupled with strength $g_0/\hbar\omega_c = 0.0563$ to a single cavity mode in resonance to $\hbar\omega_{eg}=10.746$~eV with loss $\hbar\eta=10^{-2}\hbar\omega_{eg}$ which is dressed by $N_{ensemble}$ ensemble emitters. The ensemble response was represented by a Drude-Lorentz model $V_E \chi_E(\omega)= V N_{ensemble} \cdot  \omega_p^2 /(\omega_0^2-\omega^2-i\gamma \omega), ~\hbar\omega_p=3\sqrt{10}/200~eV,~\gamma=0.1\cdot\omega_0,~\omega_0=\omega_c$.}
\label{fig:responseAlensemble_resonant}
\end{figure}

The moment we embed a macroscopic susceptibility into the dyadic Green tensor $\textbf{G}_\perp$, the spectral response of the now dressed environment will change. If the excitation energies of $\chi_E(\omega)$ and $\textbf{G}_{0}$ overlap and their interaction is non-zero, the dressed  $\textbf{G}_\perp$ will feature polaritonic excitations that are separated by the hybridization energy. Our individual molecule that hybridizes with those new quasi-particle excitations will be forced to oscillate at the excitation energies of the collective states.
Figure~\ref{fig:responseAlensemble_resonant} presents how an individual one-dimensional hydrogen contributes to the absorption of  the collective state. The larger the number of ensemble emitters $N_{ensemble}$, the smaller its participation in the bright upper and lower polaritonic states. The spectral weight of the single molecules moves into an increasing number of dark states which are located at the original excitation energy. This bright/dark state behavior represents the hallmark of collective strong coupling and is at the heart of polaritonic chemistry \cite{ebbesen2016,yuen2019polariton}. 
By embedding the ensemble-dressed cavity into the electronic structure calculations, we obtain simultaneous access to the complex electronic dynamics and its interplay with macroscopic collective states.

\begin{figure}[ht]
\includegraphics[width=1.0\columnwidth]{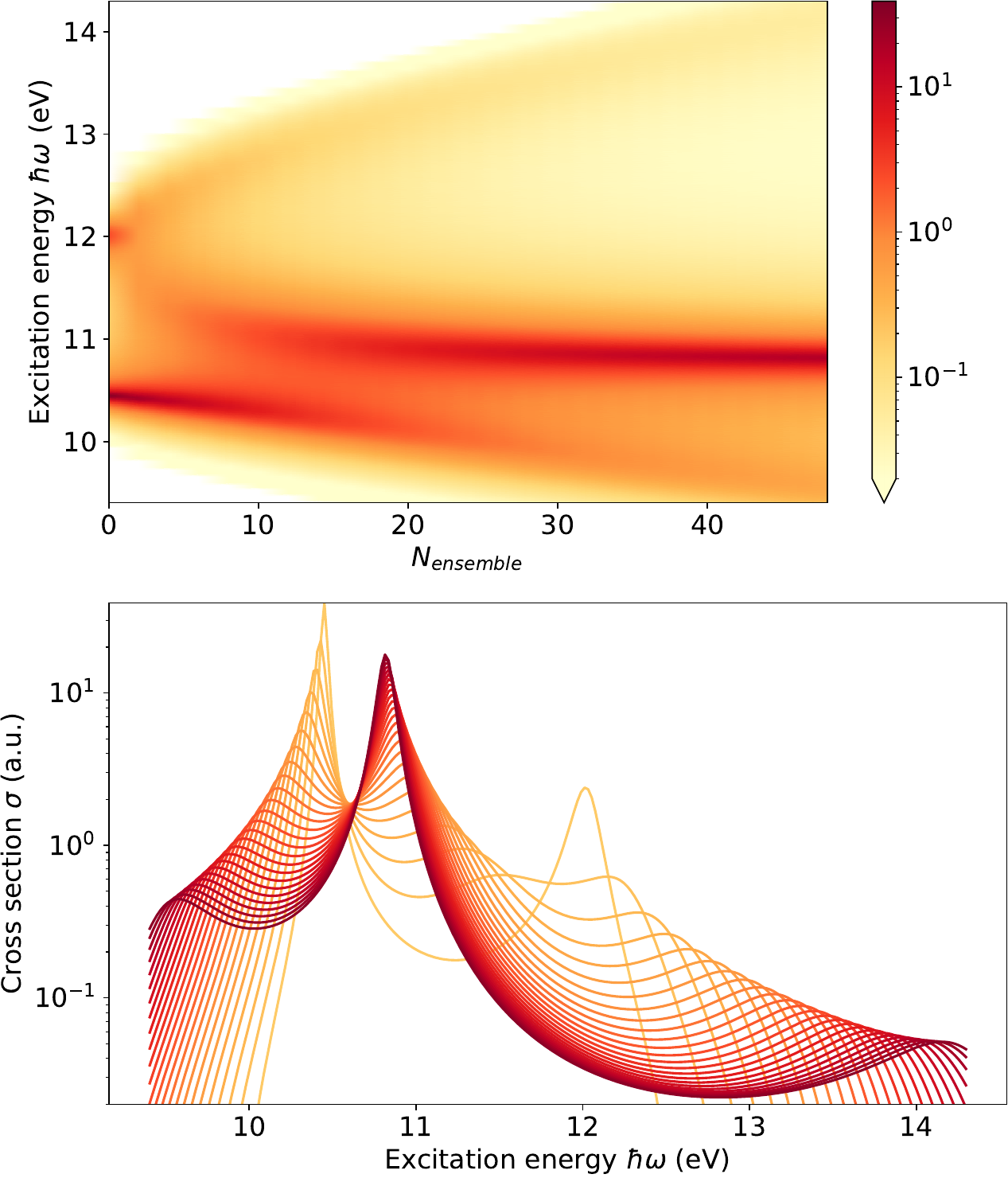}
\caption{ 
Photoabsorption cross-section $\sigma(\omega)$ of one-dimensional hydrogen coupled coupled with strength $g_0/\hbar\omega_c = 0.0539$ to a single cavity mode which is detuned at $\hbar\omega_{c}=11.7$~eV with loss $\hbar\eta=10^{-2}\hbar\omega_{eg}$ and dressed by $N_{ensemble}$ ensemble emitters. The ensemble response was represented by a Drude-Lorentz model $V_E \chi_E(\omega)= V N_{ensemble} \cdot  \omega_p^2 /(\omega_0^2-\omega^2-i\gamma \omega), ~\hbar\omega_p=3\sqrt{10}/200~eV,~\gamma=0.1\cdot\omega_0,~\omega_0=\omega_c,~V_E/V=10^{-5}$.}
\label{fig:responseAlensemble_detuned}
\end{figure}

A more complex bright-state dynamic appears if cavity, ensemble and explicit emitter are energetically detuned (see figure~\ref{fig:responseAlensemble_detuned}).

Cavity and ensemble will create the usual (large symmetric) splitting to which the detuned hydrogen contributes weakly. At larger $N_{ensemble}$, the bright ensemble states start to hybridize with the bare excitation of hydrogen, leading to upper, middle and lower polaritonic bright states. The hybridization strength between middle and lower polariton is dominated by the fundamental coupling strength between hydrogen and cavity.
Close to the avoided crossing, the contribution of hydrogen to the hybridized state can be substantially larger than the contribution of an average ensemble emitter --- at this point hydrogen and the bright ensemble state contribute equally to the full polaritonic states. 
This becomes apparent when comparing figure~\ref{fig:responseAlensemble_resonant} to figure~\ref{fig:responseAlensemble_detuned}.
Similar effects have been also observed with quantum-electrodynamical density-functional theory \cite{sidler2020polaritonic}. Appendix~\ref{app:hop} illustrates how the embedding radiation-reaction approach and the here discussed observations can be understood from the perspective of simple quantum optical models.

An important limitation of this simplified approach rests in the self-consistency of the embedding procedure and the strength of any external perturbation. We made two assumptions, first, the ensemble can be described as linear medium, i.e., external fields are weak. While we could apply the radiation-reaction potential without issue for the full ensemble at once to treat e.g. high-harmonic generation \cite{schafer2021shortcut}, i.e., $\varepsilon_r=1,~\textbf{j}=\sum_{i=1}^{N_E+1}\textbf{j}_i$ and $\textbf{G}=\textbf{G}_0$, the resulting computational cost would be prohibitive for larger systems.
The second, the cavity field exerts only weak forces on the ensemble. This limit is fulfilled in the weak and strong coupling regimes for light-matter interaction, i.e., cavity and ensemble can hybridize but the underlying matter-response experiences only minor changes. When reaching the ultra-strong coupling regime between light and matter, the underlying electronic or nuclear excitation structure will be modified considerably  \cite{schafer2018insights,haugland2020intermolecular,schafer2021making,flick2017c,latini2021ferroelectric}. We can account for those change by e.g. using the QEDFT local-density approximation proposed in \cite{schafer2021making} which is fully compatible with the here proposed radiation-reaction potential. We demonstrate such a combination in App.~\ref{app:lda} and show that, for the here selected parameter regime, the QEDFT corrections are negligible.
Furthermore, we can self-consistently update all components if extensions into stronger coupling regimes are envisioned. Imagine we couple a small set of molecules very strongly to the cavity but embed all but one into the Green tensor, substantial light-matter interaction will change the response of each molecule, which in turn affects the dressed Green tensor --- a  self-consistent cycle emerges in which $\chi,~\textbf{j}$ affect $\textbf{G}$ and the latter modifies $\chi,~\textbf{j}$. Such a generalized embedding scheme, which appears particularly appealing in the linear response regime, and its combination with the local-density approximation will be the subject of future work. The real-time propagation employed in the present work provides natural access out-of-equilibrium molecular dynamics and conveniently integrates cavity and ensemble lifetimes into the microscopic description.

Collective strong coupling experiments are safely within the strong coupling regime and typically performed in the 'dark', i.e., without or with only weak external drive \cite{garcia2021manipulating,genet2021inducing,simpkins2021mode}. The conditions of the illustrated approach are therefore satisfied. Most importantly, both limitations are only mildly limiting the dynamic of the explicitly treated molecule. It is entirely within the limits of the approximation if our explicit molecule undergoes for instance a charge reorganization or a chemical reaction --- both considered to be of major interest in QED chemistry and QED material design.

\subsection{Embedding Radiation-Reaction for Realistic Systems}

Our aspiration is to describe the dynamic of realistic molecules in collective light-matter states. For this purpose, the embedding radiation-reaction ansatz has been implemented into GPAW \cite{enkovaara2010electronic} --- a simple task as the local potential is easily combined with the TDDFT framework. 
This allows us to validate the previous conceptions for a realistic system, here chosen to be a chain of sodium dimers in H-aggregate configuration. The dimers as well as the cavity are oriented in z-direction, meaning that $ \boldsymbol\epsilon_c^T \cdot \boldsymbol\alpha(\omega) \cdot \boldsymbol\epsilon_c = \alpha_{zz}(\omega) $. The ensemble dressed dyadic for this example reads then
\begin{align}
\textbf{G}_\perp(\omega) &\approx \boldsymbol e_z \boldsymbol e_z^T \left(g^{-1}_0(\omega) - \frac{\omega^2}{c^2} \frac{N_E}{\varepsilon_0} \alpha_{zz}(\omega) \right)^{-1}.
\end{align}

\begin{figure}[ht]
\includegraphics[width=1.0\columnwidth]{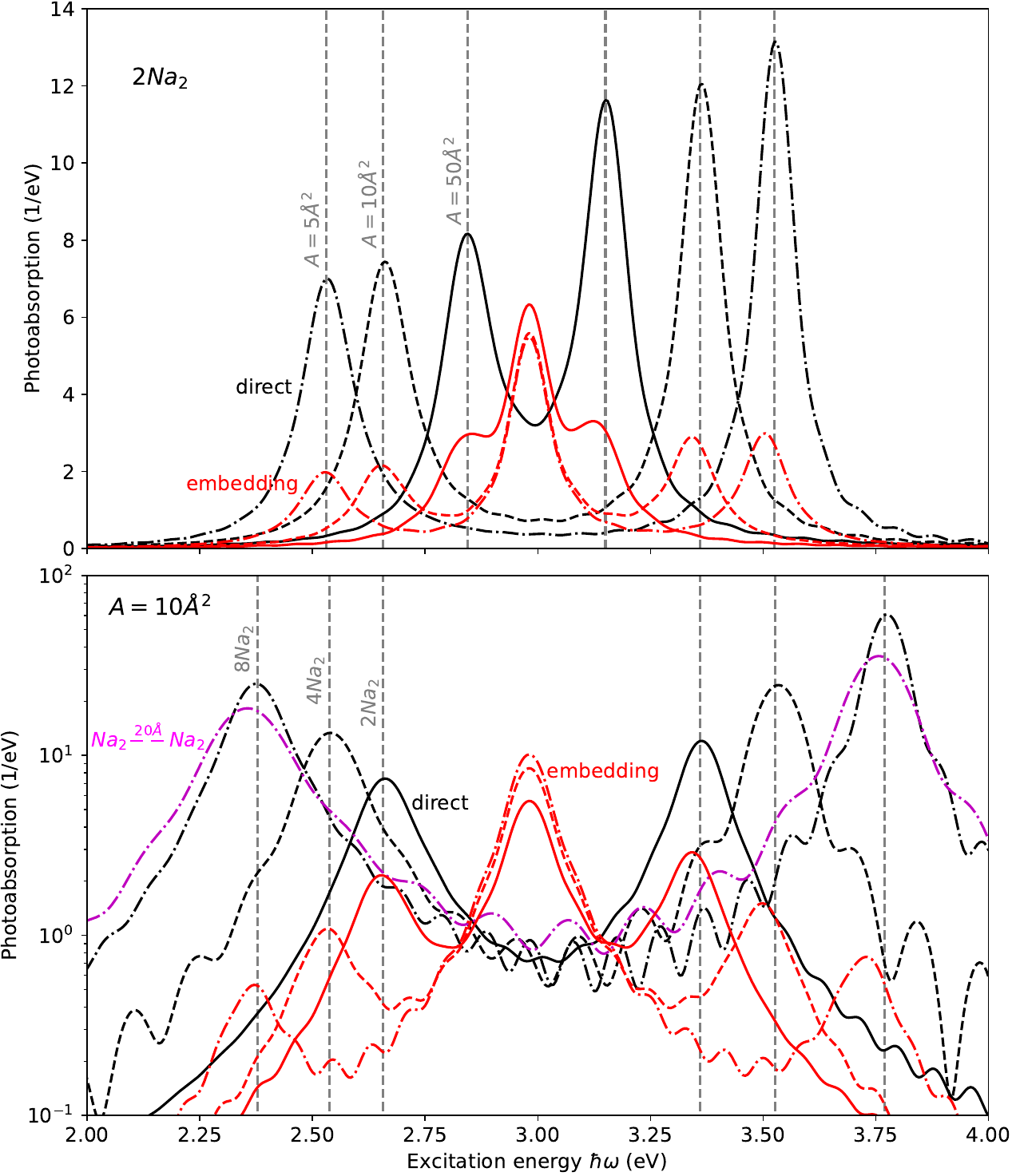}
\caption{ 
Photoabsorption of $Na_2$ chains coupled to a single cavity mode with $\hbar\omega_{c}=2.98$~eV and $\hbar\eta_c=10^{-2}\hbar\omega_{c}$. The response is calculated either directly using TDDFT+radiation-reaction for all N dimers (black) or using the embedding radiation-reaction (red) where only a single dimer is calculated explicitly. Different combinations of light-matter coupling strengths ($g \propto 1/\sqrt{A}$) and particle numbers $N$ are illustrated. Vertical gray-dashed lines serve as guide to the eye for specific polaritonic resonances.
Note that the embedding approach describes the contribution of a single dimer to the collective state, i.e., the majority of the spectral weight is located in the dark states which is not visible in the direct calculations (as it is dark). Excitation energies and spectral weight are in excellent agreement otherwise. The direct calculations exhibit a weak blue-shift that can be removed by increasing the distance between dimers from 10 \AA~ to 20 \AA~ (magenta). Its origin lies in the Coulomb-mediated dipole-dipole interactions between the chain-elements. The excitations have been artificially broadened, numerical details in appendix~\ref{app:numerics}.}
\label{fig:responsesoidum}
\end{figure}

We follow now the in section \ref{sec:collective} illustrated flow. First, we obtain the bare response of a single sodium dimer using the radiation-reaction potential \cite{schafer2021shortcut} with the cross sectional area $A=10^2$~\AA. Next, we obtain the polarizability $\alpha_{zz}(\omega)$ which defines $\textbf{G}_{\perp}(\omega)$ and with it the corresponding embedding radiation-reaction potential. Finally, a linear-response calculation of a single sodium dimer that is affected by the ensemble dressed cavity is performed. Figure~\ref{fig:responsesoidum} sets this embedding approach (red) in relation with the direct response of N sodium dimers (black) coupled to the cavity mode. Even when varying the number of sodium dimers and the cavity volume ($g\propto 1/\sqrt{V}$ where $V=A \pi c / \omega_c$), the embedding approach accurately recovers the direct calculations. The most notable difference is of course that the embedding approach describes the response of a single dimer, i.e., most of the spectral density is located in the dark state which is not visible in the direct N-dimers+cavity calculations. The polaritonic excitations in the embedding calculations show however the expected $1/N$ trend, e.g., $(2~Na_2)/(4~Na_2)\approx 2$, as well as the correct asymmetry between upper and lower polariton. Weak Coulomb mediated dipole-dipole interactions result in a slight blue-shift, as expected from H-aggregates \cite{schafer2021shortcut}. Increasing the distance between the dimers to 20~\AA~alleviates this effect and we find excellent agreement between direct and embedding approach. As described earlier, Coulomb mediated couplings can be included in the longitudinal component of the embedding radiation-reaction ansatz.

Using the embedding radiation-reaction ansatz is substantially cheaper than calculating the full dimer chain explicitly. While the cost of the latter grows quickly beyond the manageable, our embedding approach retains the cost of a single-dimer calculation. This computational advantage opens up the possibility to investigate far more complex systems, incorporating the experimentally relevant realizations of collective strong coupling which will be the subject of future work. 
While we have limited ourselves thus far on the spectral properties of collective strong coupling, we will demonstrate in the following how chemistry itself is altered. 

\subsection{Modifying chemistry with collective strong coupling}

We concluded previously that the participation of a single molecule to the collective state will usually decrease with increasing number of molecules, as expected from quantum-optical models. Nevertheless, a single molecules that exhibits a detuned excitation will contribute substantially stronger at the avoided crossing compared to a 'regular' molecule of the resonant ensemble (e.g. figure~\ref{fig:responseAlensemble_detuned}).
During a chemical reaction, usually triggered by thermal fluctuations, structural changes of the molecule will alter its vibrational spectrum. In between reactant and product state, the molecule will spend time in intermediate configurations which will likely feature different vibrational energies and oscillator strengths. The closer this detuned vibration is to the collective polaritonic states, the stronger its contribution will out-weight the individual contribution of an ensemble molecule.
While the reactant configuration will therefore play a negligible role in the polaritonic state, during the reaction our individual molecule might play a dominant role. Even if the number of dark-states might be large, the bright states can play a dominant role in this scenario.
Quantitative estimates depend certainly on the specific characteristics of the molecule and its reaction. As we will see in the following, especially relevant is its vibrational structure and shifts thereof in intermediate configurations as well as reaction energetics, Stark-shifts and kinetics in relation to light-matter exchange times.

\begin{figure}[h]
\includegraphics[width=1.0\columnwidth]{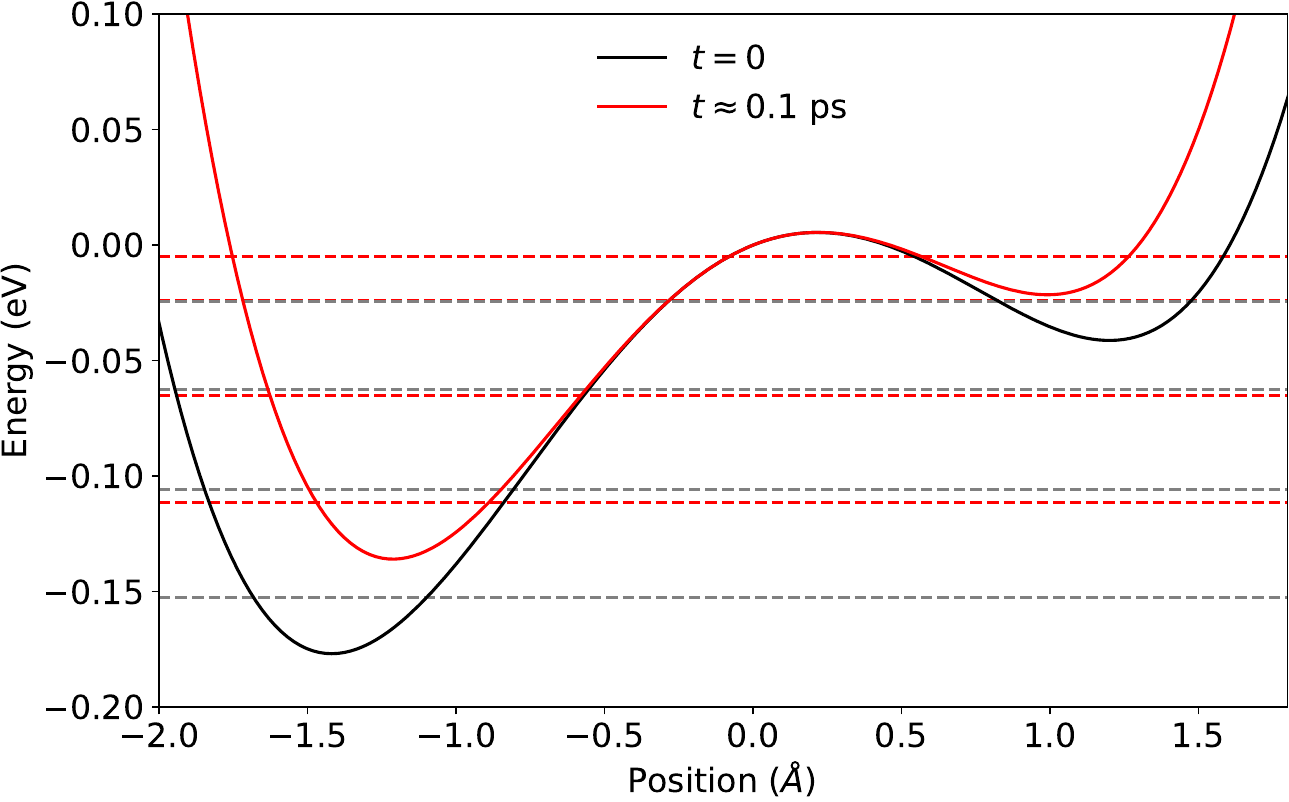}
\caption{ Potential energy surface before (black) and after (red) the deformation. Dashed horizontal lines indicate eigenvalues of the respective static single-proton Schr\"odinger equation.}
\label{fig:reactivity_potential}
\end{figure}

Let us illustrate this concept for a simple proton-tunneling model that is designed to represent ground-state reactions. A tilted double well potential $v(x,t) = x\cdot 10^{-3}-x^2 \cdot 1.25\cdot 10^{-3} + x^4\cdot (1 + \frac{0.4}{1+e^{-(t-60fs.)/10fs}})\cdot 10^{-4}~H$, shown in figure~\ref{fig:reactivity_potential}, is slightly deformed over a time-frame of approximately $t_p\approx 0.1$~ps. Such a deformation results in a weak excitation of the proton wavefunction and leads to a finite probability to overcome the barrier from the left to the right well. The amount of nuclear density that overcomes the barrier will be considered as an indication for the reactivity in our model. We start by setting cavity and ensemble excitation in resonance.

\begin{figure}[h!]
\includegraphics[width=1.0\columnwidth]{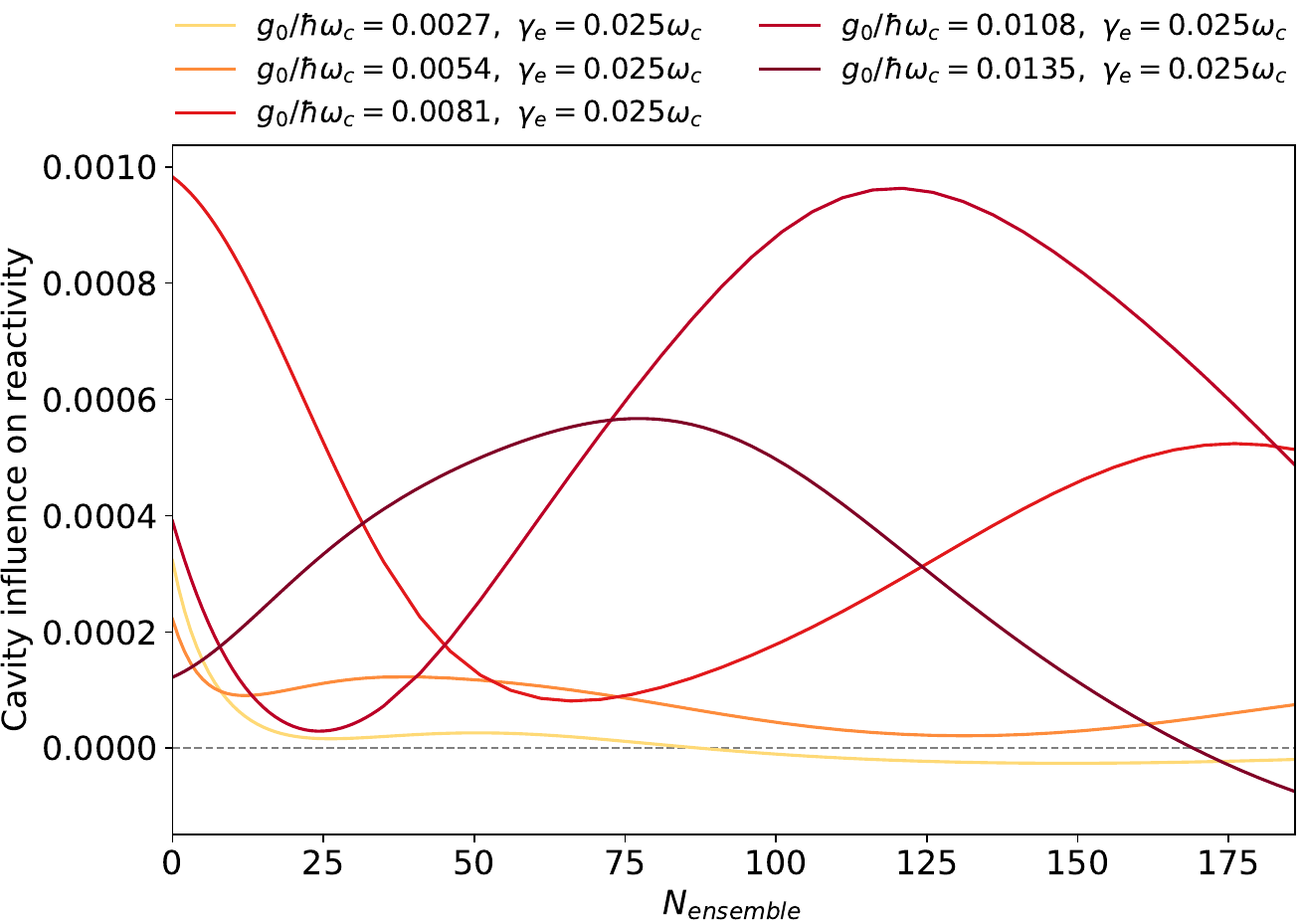}
\includegraphics[width=1.0\columnwidth]{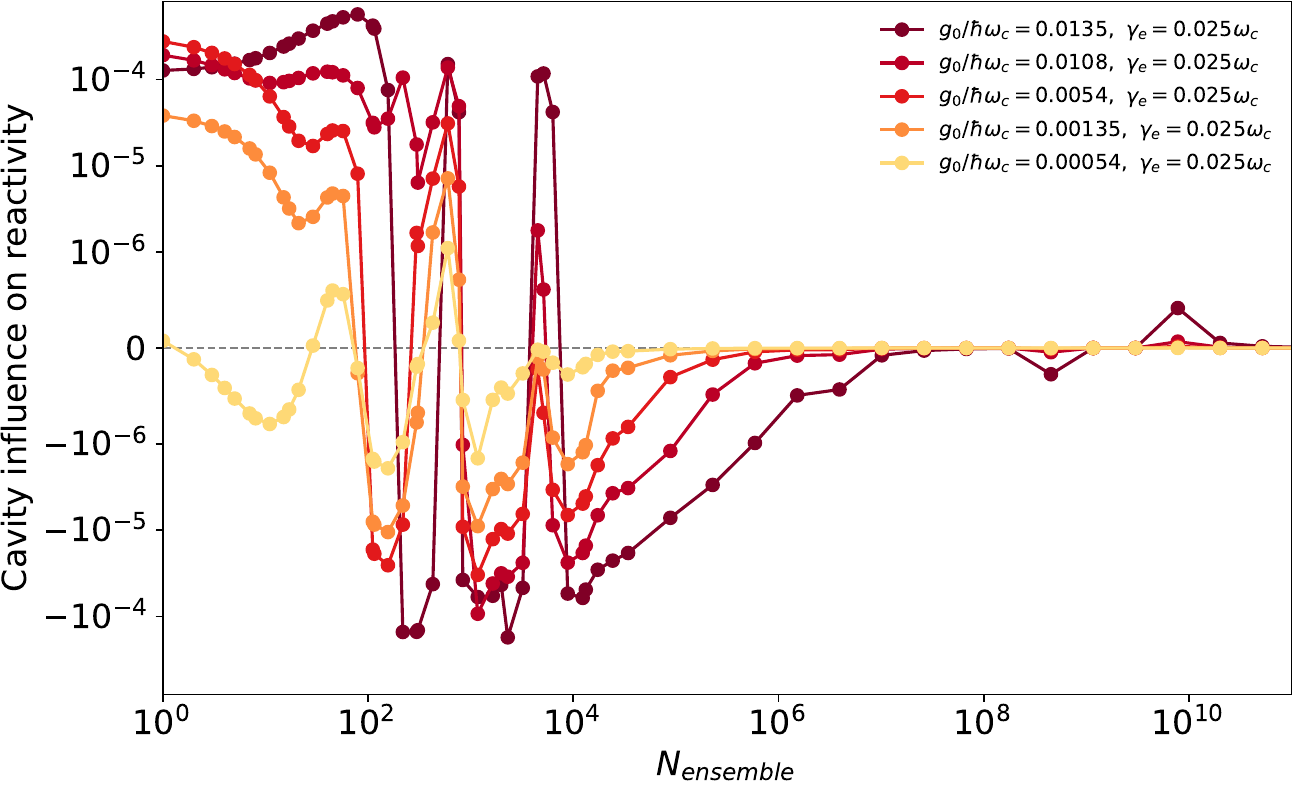}
\caption{ 
Cavity influence on reactivity $CR = \int_0^T \frac{dt}{T} [1-(\int_{-\infty}^{0} dx n(x,t) - \int_0^{\infty} dx n(x,t))] - CR(g_0=0)$ for increasing number of ensemble emitters for various coupling strengths plotted on linear (top) and logarithmic (bottom) scales. The dependence of the cavity influence on the number of ensemble molecules $N_{ensemble}$ is non-trivial and oscillates between facilitating and inhibit the reaction. The cavity-influence on the chemical reactivity is far from a simple $1/N_{ensemble}$ behavior. Especially intriguing, for small fundamental coupling strength, the inhibiting effect of collective strong coupling increases with increasing ensemble size. Up to $N_{ensemble} \approx 10^4$, the absolute influence on the reaction remains comparably stable, followed by a slow decay when reaching the thermodynamic limit. Higher fundamental coupling strengths ($g_0/\hbar\omega_c = 0.0135,~0.0108$) feature even in the thermodynamic limit ($N_{ensemble} \gg 10^8$) clearly visible resonant effects. It should be noted, however, that the collective hybridization strength reaches substantial values in this case (see e.g. figure \ref{fig:reactivity_greensfunction_largeN_imagG}) and the underlying approximations are expected to limit the domain for reliable predictions to a combination of smaller fundamental couplings and ($N_{ensemble} \ll 10^6$).
We use the in figure~\ref{fig:responseAlensemble_resonant} introduced Drude-Lorentz model with $\hbar\omega_p=6.387 \cdot 10^{-4}~eV$. The first excitations are aligned at t=0 $\hbar\omega_c=0.046721~eV=\hbar\omega_E=\hbar\omega_{0-1}$, the cavity decay rate is fixed to $\eta=10^{-4}\omega_c$.
}
\label{fig:Nensemblelarge}
\end{figure}

Figure~\ref{fig:Nensemblelarge} shows the influence of collective light-matter coupling on the accumulated proton-tunneling. 
The influence of the ensemble on the reaction depends largely on the single-particle coupling strength to our impurity molecule and even for $N_{ensemble}=0$ is the reactivity already noticeably increased. Smaller coupling values (e.g. $g_0/\hbar\omega_c=0.0027$) show the tendency to decrease this effect with increasing ensemble size. However, even this comparably stringent trend features oscillations which can result in an inhibition of the reaction for larger $N_{ensemble}$. 
Increasing the fundamental light-matter coupling does not only shift the resulting curve to smaller $N_{ensemble}$ values but furthermore intensifies the overall trend. In the extreme case of $g_0/\hbar\omega_c=0.0135$, we observe initially an increase in catalysis followed by an inhibiting effect for large $N_{ensemble}$. Both, catalyzing \cite{lather2019cavity} and inhibiting effect have been observed in experiments using either vibrational \cite{thomas2016,thomas2019tilting} or electronic strong coupling \cite{hutchison2012,munkhbat2018suppression}.

\begin{figure}[h]
\includegraphics[width=1.0\columnwidth]{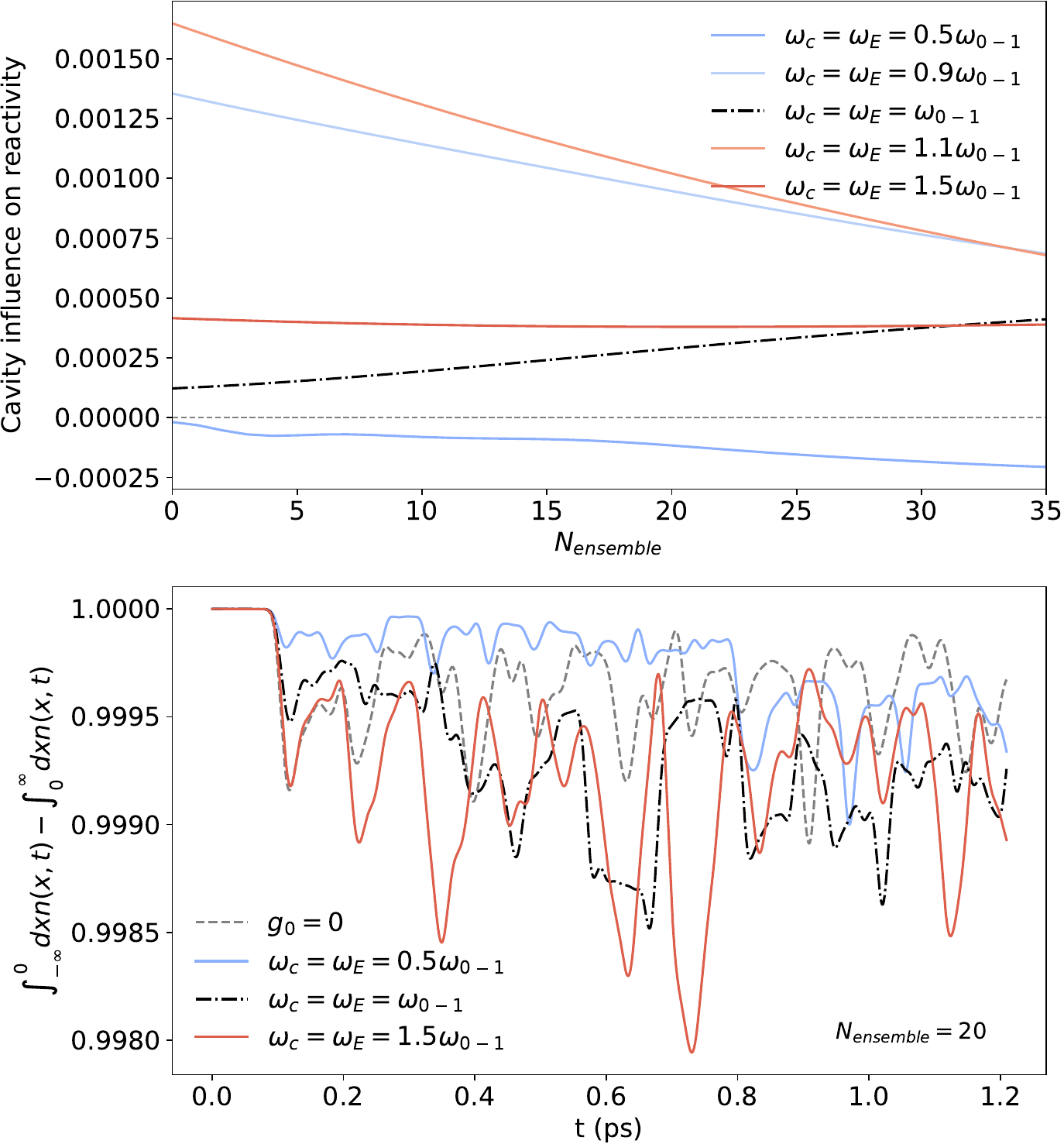}
\caption{ 
Top, cavity influence on reactivity $CR = \int_0^T \frac{dt}{T} [1-(\int_{-\infty}^{0} dx n(x,t) - \int_0^{\infty} dx n(x,t))] - CR(g_0=0)$ for increasing number of ensemble emitters for different energetic alignments between cavity/ensemble frequency and bare excitation frequency of our impurity molecule.
Bottom, proton-tunneling $\int_{-\infty}^{0} dx n(x,t) - \int_0^{\infty} dx n(x,t))$ for $N_{ensemble}$. The complex excitation-structure leads to a non-trivial influence on the proton-tunneling. In contrast to the other parameters, the smallest frequency tends to inhibit the proton-tunneling process. We fixed the cavity volume $V$ in all calculations to the value of the resonant alignment. It is apparent from the behavior with increasing ensemble number $N_{ensemble}$, that the specific effect of the cavity on the chemical reactivity is sensitive to the energetic alignment between cavity, ensemble and reacting molecule as well as the hybridization between those systems.
}
\label{fig:reactivity_cavitydetuning}
\end{figure}

Figure~\ref{fig:reactivity_cavitydetuning} (top) illustrates that the influence of collective light-matter coupling is sensitive to the energetic alignment between ensemble, cavity mode and reacting system. The associated charge-transfer dynamic is shown in figure~\ref{fig:reactivity_cavitydetuning} (bottom).
Such a complex behavior for our simple model demonstrates that the influence of collective strong coupling on chemical reactivity can be highly situational --- a strong argument for the importance of \textit{ab initio} approaches.

\begin{figure}[h]
\includegraphics[width=1.0\columnwidth]{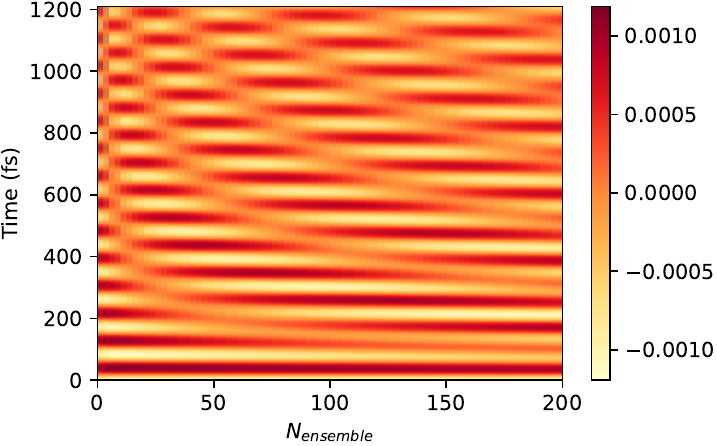}
\caption{ 
Time-domain convolution-argument $\mathcal{F}_t^{-1} i \omega G(\omega)$ for $g_0/\hbar\omega_c=0.0135$, $\gamma_e=0.025 \omega_c,~\eta = 10^{-4} \omega_c$. Only for sizable hybridization strength can we observe the Rabi-period on the time-scale of the reaction-time.
}
\label{fig:reactivity_greensfunction_largeN}
\end{figure}

The dynamic that gives rise to this behavior emerges from the interplay of single-particle and many-body excitations in the correlated system and can be best understood by investigating the Green tensor of the ensemble-dressed cavity in figure~\ref{fig:reactivity_greensfunction_largeN} (here $g_0/\hbar\omega_c=0.0135$).
Increasing Rabi-splitting ($\propto \sqrt{N_{ensemble}}$) results in a shortening of the beating period observed in $\mathcal{F}^{-1}_t i\omega G(\omega)$. As we limited the reaction-time to approximately 1.2~ps, the qualitative influence of the ensemble will depend on how strong $\mathcal{F}^{-1}_t i\omega G(\omega)$ is noticeably altered on this time-scale. This defines a minimal hybridization strength of $\textbf{G}(\omega)$ for a given reaction-speed $1/\omega_{Rabi} \sim T_{Rabi} < \tau_{reaction} $ --- a quick reaction demands a large hybridization. How a chemical reaction will be influenced by collective strong coupling will therefore depend not only on the energetic alignment between reactant, product, solvent and cavity but also its hybridization strength in relation to the reaction time (see also reference~\cite{schafer2021shining}).

Quicker deformation of the potential-energy surface injects more energy into the system, leading to a larger reactivity and an enhanced cavity influence. Moreover, quicker ensemble decoherence leads to broad spectral response which tends to flatten the influence of collective strong coupling on the proton tunneling. Both effects are shown in appendix~\ref{app:loss_pspeed}.

An important realization is now that the mere necessity of the existence of collective polaritonic states defines a resonant condition. In other words, for the cavity to have an effect dependent on $N_{ensemble}$, collective strong coupling has to be present and an avoided crossing and energy-redistribution between shifted intermediate configuration and bare molecules has to be possible. Utilizing collective strong-coupling for chemistry along the here presented mechanism would therefore be limited to a subset of reactions with the correct intermediate-state dynamics. While experimental efforts strive to extend the realm of QED chemistry, among many experimental attempts only few reactions (all with similar molecular characteristics) have been successfully identified. Such a limited applicability of polaritonic chemistry would agree with the here suggested complex pre-requisites and calls for continued effort to identify further reactions in order to allow a conclusive understanding. This effort should consider the question to which extend for instance solvation can influence the number of effectively coupled molecules in collective states.

Identifying the specific influence of collective strong coupling on a chemical reaction remains an open and theoretically challenging problem. While recent work \cite{schafer2021shining,li2021collective,li2021cavity} showed promising progress, much remains to be investigated and a conclusive understanding will likely demand a collaborative effort involving experiment, \textit{ab initio} theory and simplified models.

\section{Conclusion}

By embedding an ensemble of molecules into the recently proposed radiation-reaction potential \cite{schafer2021shortcut}, we have been able to tackle the conundrum of a first-principles description of the material structure combined with the collective light-matter interaction involving a large number of emitters. The embedding ansatz is trivial to implement into existing time-dependent density-functional theory libraries as it adds a simple local potential. This allowed us to treat collective strong coupling featuring arbitrarily large number of emitters with marginal additional computational cost compared to ordinary real-time TDDFT calculations. We discussed a hierarchy of possible approximations and their associated limitations. Their utilization was first demonstrated for the linear response of a single molecule embedded in the electromagnetic environment of a cavity dressed by an ensemble of emitters. 
Integrating the embedding radiation-reaction potential into GPAW allowed us to tackle ensembles of realistic molecules from first principles. Lastly, by tuning the cavity in the infrared regime, we demonstrated that the modification of proton-tunneling reactions via collective strong coupling can posses a non-trivial dependence on the number of ensemble molecules.

This work paves the way to tackle new regimes of theoretical chemistry, accounting for the self-consistent interaction between light with macroscopic systems while focusing on the microscopic dynamic of single subsystems --- entirely from first principles. Possible extensions include a consistent treatment of near-field effects that are relevant for solvation, plasmonic environments and J/H-aggregates.
Equipped with this tool-set, describing consistently the seminal experimental work in polaritonic chemistry \cite{ebbesen2016,garcia2021manipulating,simpkins2021mode} moves within reach. This allows us to relax previous simplifications to only single or few molecules with artificially increased coupling strength \cite{schafer2021shining,luk2017multiscale,sidler2020polaritonic,bonini2021ab}. Already our here illustrated simple proton-tunneling model exhibited a nontrivial behavior with the number of emitters. We can expect that complex multi-step chemical reactions will further complicate this trend.
The seamless combination of embedding radiation-reaction and quantum-electrodynamical density-functional theory, especially the recently developed local-density approximation \cite{schafer2021making}, ensures that quantum-corrections of the light-matter interaction can be properly accounted for if necessary. 
Collective strong coupling for chemistry and material design represent timely theoretical challenges to which the here presented embedding radiation-reaction potential provides an important and novel pathway.

\begin{acknowledgments}
I thank G\"oran Johansson for interesting discussions as well as Jakub Fojt for assistance with GPAW. This work was supported by the Swedish Research Council (VR) through Grant No. 2016-06059 and the computational resources provided by
the Swedish National Infrastructure for Computing at Chalmers Centre for Computational Science and Engineering partially funded by the Swedish Research Council through grant agreement no. 2018-05973.
\end{acknowledgments}

\appendix

\section{Illustration of the embedding at the Hopfield model}\label{app:hop}
We illustrate here how the embedding procedure can be understood from the perspective of the quantum-optical Hopfield model for hybridizing modes. This widely used and strongly simplified model assumes the dipolar approximation, the rotating-wave approximation and emitters are approximated as bare 2-level systems. Expanding the corresponding Hamiltonian $\hat{H}_{sH} = \omega_c\hat{a}^\dagger\hat{a} + \sum_i^N \omega_i \sigma_{z,i} + \sum_i^{N+1} g_i (\hat{a} \sigma^+_{i} + \hat{a}^\dagger \sigma^-_{i})$ into the excitation basis leads to a matrix equation that is commonly fitted in order to analyze experimental data.

The interaction between an ensemble of collectively interacting modes, describing a large set of ensemble molecules, a selected single molecule, and a cavity mode, will be given under the above approximations by the simplified Hamiltonian-matrix
\begin{align*}
\begin{pmatrix}
\omega_{c} & g & \cdots & g_m \\
g & \omega_{E,1} & 0 & 0\\
g & 0 & \ddots & 0 \\
g_m  & \cdots & 0 & \omega_m
\end{pmatrix}
\end{align*}
for which the $N$ ensemble molecules posses the identical excitation energy $\omega_{E,1}=\omega_{E,N}$ and light-matter coupling $g$. In this case, arrow-head structure allows us to remove all those degenerate eigenvalues from the matrix by separating the problem into symmetric (also known as bright) and antisymmetric (also known as dark) sub-blocks.

For example, the simplest case of this block-diagonalization is given for 2 emitters coupled to one mode. The Hamiltonian matrix  
\begin{align*}
\begin{pmatrix}
\omega_{c} & g & g \\
g & \omega_{E} & 0\\
g & 0 & \omega_{E} \\
\end{pmatrix}
\end{align*}
can be separated via $\textbf{S}^T \textbf{H} \textbf{S}$ where
\begin{align*}
\textbf{S} =
\begin{pmatrix}
1 & 0 & 0 \\
0 & 1/\sqrt{2} & -1/\sqrt{2}\\
0 & 1/\sqrt{2} & 1/\sqrt{2} \\
\end{pmatrix}
\end{align*}
into block-structure
\begin{align*}
\begin{pmatrix}
\omega_{c} & \sqrt{2}g & 0 \\
\sqrt{2}g & \omega_{E} & 0\\
0 & 0 & \omega_{E} \\
\end{pmatrix}
\end{align*}
where the cavity couples now only to the symmetric/bright state with amplified strength $\sqrt{2}g$ while the antisymmetric/dark state is decoupled from the cavity. The same can be achieved by noticing that for identical coupling and excitation energy the sum of excitation operators $ \sum_i^{N+1} g_i (\hat{a} \sigma^+_{i} + \hat{a}^\dagger \sigma^-_{i}) $ can be simplified to collective excitation operators $S^{\pm} = \sum_i^N \sigma_i^{\pm}$, also called Dicke operators.

\begin{figure}[t]
\includegraphics[width=1.0\columnwidth]{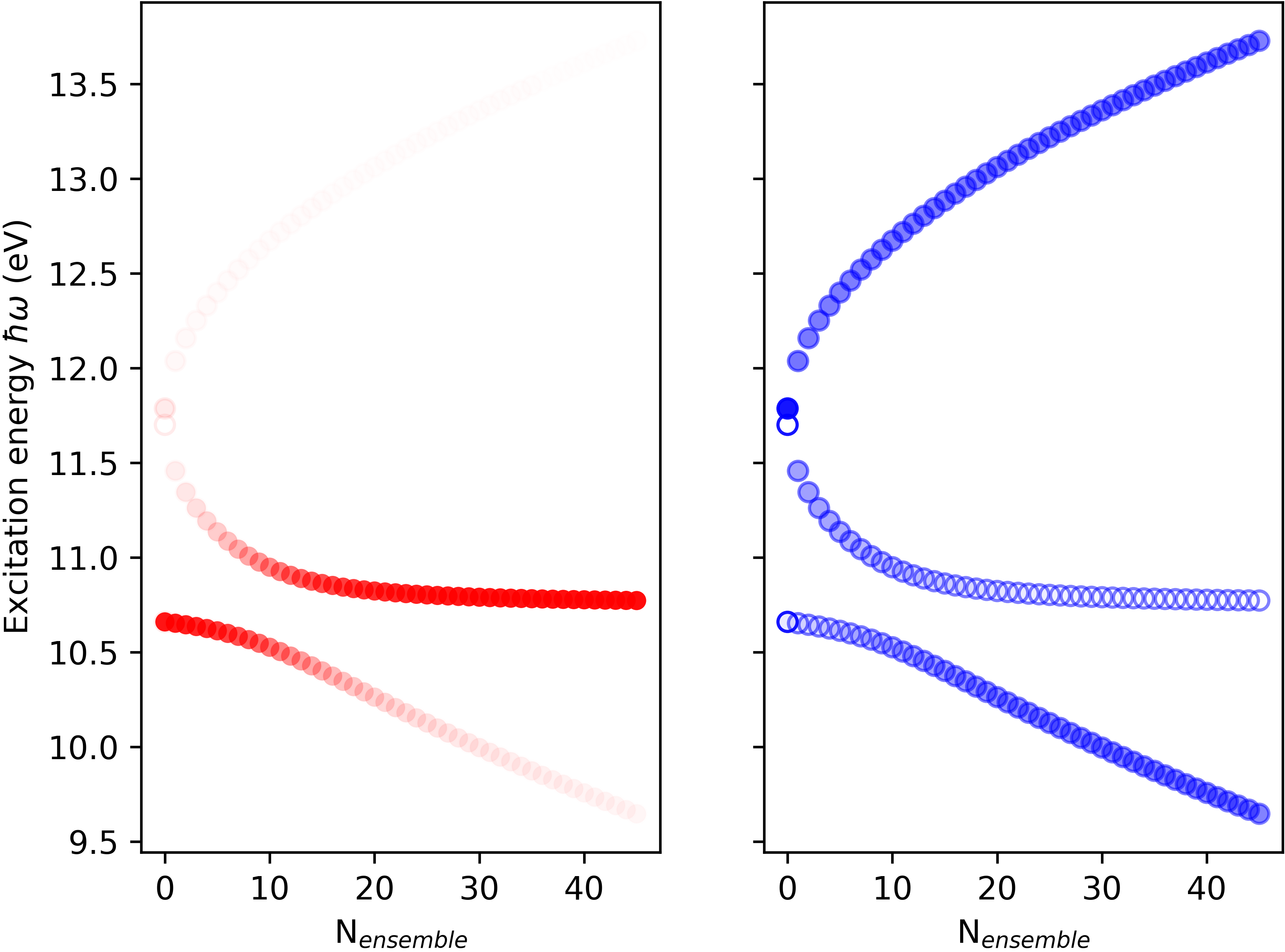}
\caption{ Simple Hopfield model for the interaction of an ensemble of identical molecules coupled to a single cavity mode which again is coupled to a single detuned molecule. Left, contribution of the single molecule to the collective states. Right, photonic contribution. Parameters $\hbar\omega_c=\hbar\omega_E=11.7~eV,~\hbar\omega_m=10.746~eV$, $g=g_m=0.011$.}
\label{fig:hopfielddetuned}
\end{figure}

We reduce ourselves now to the bright-state sub-manifold described by
\begin{align*}
\begin{pmatrix}
\omega_{c} & g\sqrt{N} & g_m \\
g\sqrt{N} & \omega_{E} & 0\\
g_m  & 0 & \omega_m
\end{pmatrix} = \begin{pmatrix}
\omega_{E} & g\sqrt{N} & 0 \\
g\sqrt{N} & \omega_{c} & g_m\\
0  & g_m & \omega_m
\end{pmatrix}
\end{align*}
with the excitation structure illustrated in figure \ref{fig:hopfielddetuned}.

At the moment, the ensemble and the single molecule couple still to the bare mode.
Let us assume $\omega_c=\omega_E=1$ for simplicity. The upper-left 2-by-2 block possesses the eigenvalues $\omega_{\pm}=1\pm g\sqrt{N}$ and eigenvectors $\textbf{v}_{\pm}^T=(\pm 1/\sqrt{2}, 1/\sqrt{2})$. The transformation $\textbf{S}^{-1}\textbf{H}\textbf{S}$ with $\textbf{S} = ((\textbf{v}_1,0)^T, (\textbf{v}_2,0)^T, (0,0,1)^T)$ leads to
\begin{align*}
\begin{pmatrix}
\omega_+ & 0 & g_m/\sqrt{2} \\
0 & \omega_- & g_m/\sqrt{2}\\
g_m/\sqrt{2}  & g_m/\sqrt{2} & \omega_m
\end{pmatrix}~.
\end{align*}
and the single molecule couples now to the collective bright states (upper and lower polariton $\omega_{\pm}$). The embedding procedure via the ensemble-dressed Green tensor is conceptually similar. The dressed $\textbf{G}$ has poles at the bright ensemble+environment eigenvalues with adjusted strength, dark states are not present as they do not provide any spectral intensity from the perspective of the cavity. The microscopic material described via the Schr\"odinger equation is then affected by radiation-reaction forces which are maximal at the polaritonic resonances. The here introduced embedding scheme can be seen as a generalization of the simplified Hopfield model to realistic systems. This generalization allows us to investigate time-dependent local phenomena for realistic systems, i.e., it provides access to chemical reactions, charge-transfer processes and many other features while retaining the full collective interaction. Clearly, the underlying assumptions that the majority of the ensemble is perfectly identical and responds only weak/linear are strong limitations to the embedding approach. More complex systems will demand a more sophisticated separation of the different embedded species, the specific path chosen here is designed to address the conditions for collective strong-coupling in polaritonic chemistry.

\section{Ultra-strong coupling effects - combining radiation-reaction with QEDFT-LDA}\label{app:lda}

Every light-matter coupling strength greater than zero will alter the electronic and nuclear structure, reorganizing charge and breaking symmetries \cite{schafer2018insights}. Typically, those effects are referred to as ultra-strong coupling features and their second-order scaling $\mathcal{O}(g_0^2)$, compared to the $\mathcal{O}(g_0^1)$ scaling of polariton hybridization, renders them of minor relevance for the vast majority of experimental realizations. 
Exceptions are located in the domain of plasmonics \cite{baranov2019ultrastrong} and circuit-QED \cite{yoshihara2017superconducting}, the latter allows even coupling strengths on the order of the excitation energy $g/\hbar\omega \approx 1$. 
Quantum-electrodynamical density-functional theory (QEDFT) allows to consider such quantum-corrections also for realistic systems. We employ here the recently proposed photon-exchange local-density approximation (pxLDA) \cite{schafer2021making} in combination with the radiation-reaction potential, an easy task as the latter accounts for the classical Maxwell-components and the pxLDA potential provides access to quantum-fluctuations. 

\begin{figure}[h]
\includegraphics[width=1.0\columnwidth]{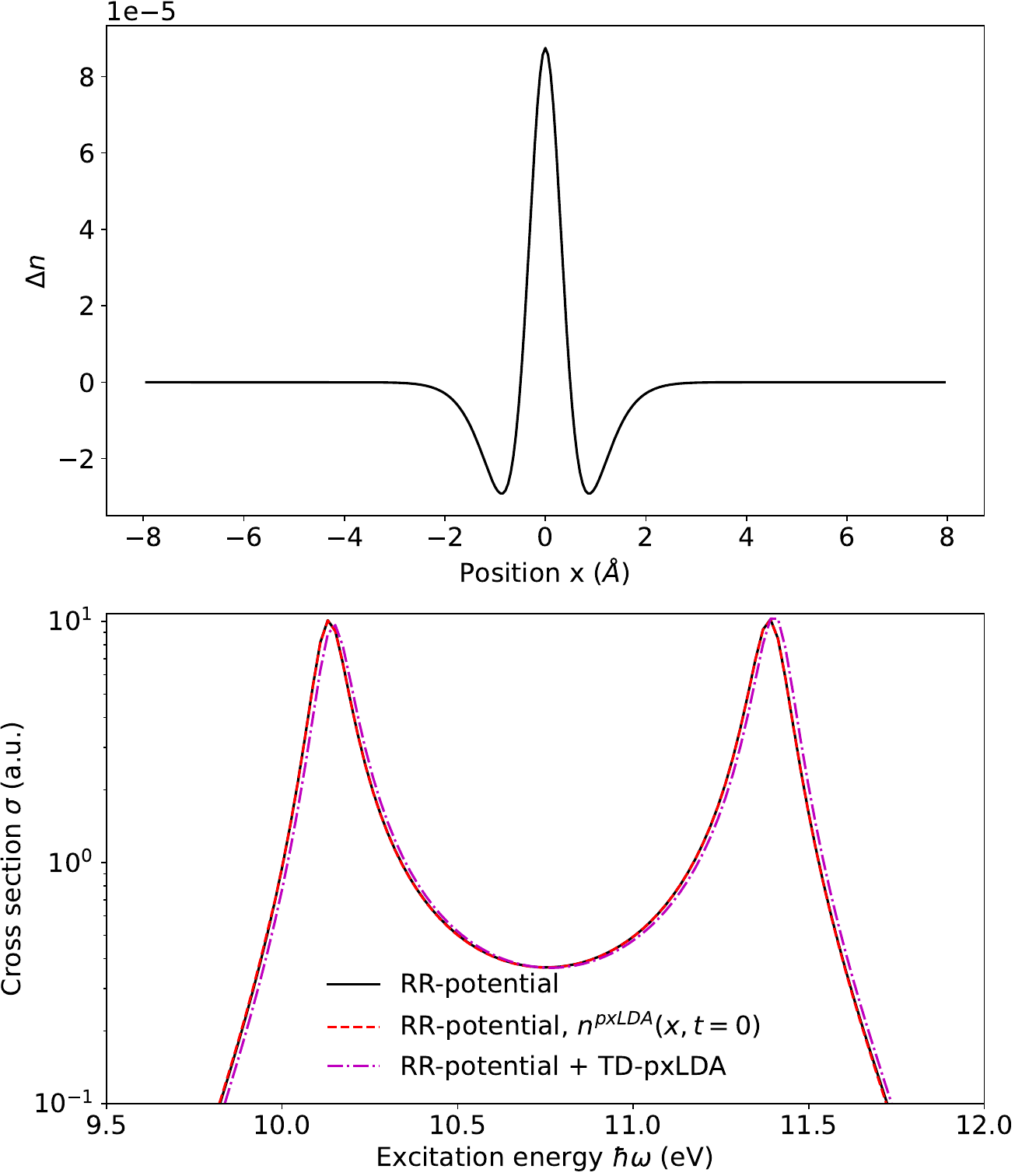}
\caption{ 
(Top) Ground-state electronic density renormalization $\Delta n(x) = n^{g_0/\hbar\omega=0.0563}(x)-n^{g_0=0}(x)$ using the pxLDA potential \cite{schafer2021making} for the system investigated in figure~\ref{fig:responseAlensemble_resonant} with $N_{ensemble}=0$.
(Bottom) Photoabsorption cross-section $\sigma(\omega)$ excluding the adiabatic pxLDA functional (black solid), accounting only for the quantum-effects on the initial-state density (red dashed), and accounting for quantum-corrections also during the time-propagation with the help of the adiabatic pxLDA approximation (magenta dashed-dotted).
}
\label{fig:complda}
\end{figure}

Figure~\ref{fig:complda} (top) exemplifies the renormalization of the electronic ground-state density for the in figure~\ref{fig:responseAlensemble_resonant} performed calculation with $N_{ensemble}=0$. The effect is minute for the chosen couplings which transfers into a negligible spectral shift (figure~\ref{fig:complda} (bottom), red dashed) based on this change of the initial-state. Usage of the adiabatic pxLDA potential during the time-propagation (magenta dashed-dotted) provides a very weak blue-shift. Consequently, the pxLDA is overall of minor relevance in our calculations and we will omit quantum-effects in the rest of this work. Note however, that such a conclusion is system and parameter specific and adding pxLDA or comparable corrections should be considered for each realization individually.

\section{Influence of loss and perturbation-speed}\label{app:loss_pspeed}

\begin{figure}[h]
\includegraphics[width=1.0\columnwidth]{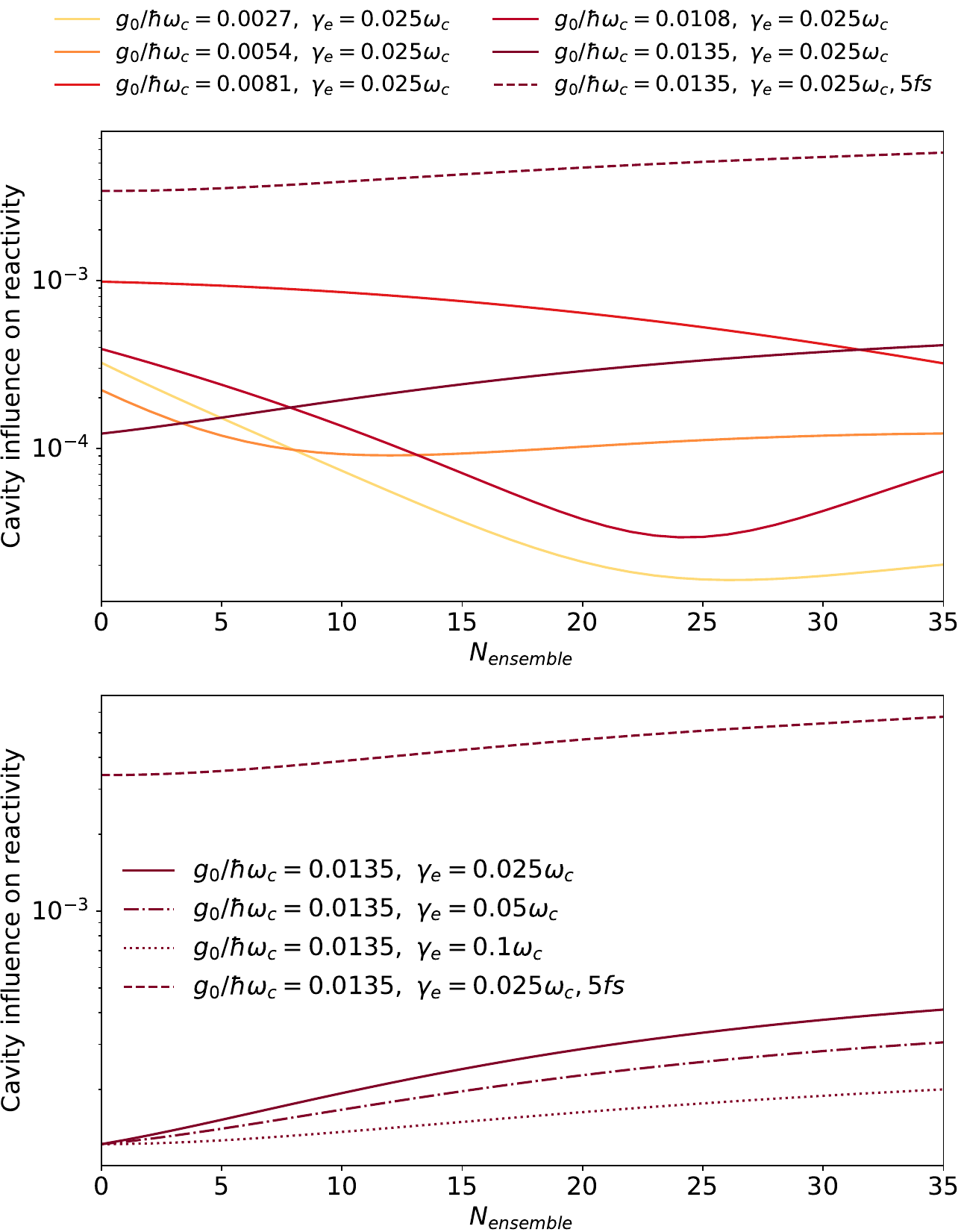}
\caption{ 
Cavity influence on reactivity $CR = \int_0^T \frac{dt}{T} [1-(\int_{-\infty}^{0} dx n(x,t) - \int_0^{\infty} dx n(x,t))] - CR(g_0=0)$ for increasing number of ensemble emitters. Top, CR for varying light-matter coupling-strength between single perturbed molecule and cavity  (the first excitations are aligned at t=0 $\hbar\omega_c=0.046721~eV=\hbar\omega_E=\hbar\omega_{0-1}$, cavity decay rate fixed to $\eta=10^{-4}\omega_c$). The dependence of the cavity influence on the number of ensemble molecules $N_{ensemble}$ is non-trivial, i.e., does not simply decrease with $1/N$ as often believed. A critical feature is the appearance of beating signatures in $G(t)$ (see figure~\ref{fig:reactivity_greensfunction_largeN}).
Shorter perturbation times (dashed, $v(x,t) = x\cdot 10^{-3}-x^2 \cdot 1.25\cdot 10^{-3} + x^4\cdot (1 + 0.4/[1+e^{-(t-5fs.)/fs}])\cdot 10^{-4}~H$) inject more energy into the system, leading to a larger reactivity and an enhanced cavity influence.
Bottom, CR for different ensemble emitter-linewidths  (solid, dashed-dotted, dotted) as well as for a shorter deformation period (dashed). Quicker ensemble decoherence leads to broad spectral response which tends to flatten the influence of collective strong coupling on the proton tunneling.
}
\label{fig:reactivity_chparity_interesting}
\end{figure}

Figure~\ref{fig:reactivity_chparity_interesting} shows the with figure~\ref{fig:Nensemblelarge} consistent calculations compared to calculations with quicker deformation (5~fs instead of 0.1~ps) or larger ensemble-loss (bottom).
Quicker deformation of the potential deposits more energy in the system, resulting in overall larger reactivity and larger absolute cavity influence.
Larger loss-rates smooth-out or flatten the cavity-influence but remain qualitatively similar to the values chosen in the main text.

\begin{figure}[h]
\includegraphics[width=1.0\columnwidth]{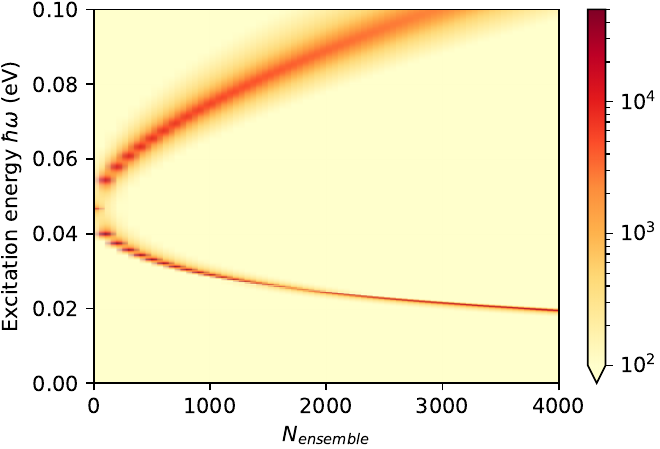}
\caption{ Excitation spectrum $\vert\Im G(\omega)\vert$ of ensemble+cavity for $g_0/\hbar\omega_c=0.0135$, $\gamma_e=0.025 \omega_c,~\eta = 10^{-4} \omega_c$.
}
\label{fig:reactivity_greensfunction_largeN_imagG}
\end{figure}

Figure~\ref{fig:reactivity_greensfunction_largeN_imagG} presents $\vert\Im G(\omega)\vert$ consistent with figure~\ref{fig:reactivity_greensfunction_largeN}.

\section{Numerical details}\label{app:numerics}

The radiation-reaction potential and the according convolution have been implemented into a small python3 library and the open-source code GPAW. The time-evolution can be efficiently calculated by finite-difference approximations.
The implementation has been checked for consistency against pre-existing Maxwell-QEDFT calculations \cite{schafer2021making}.

Close to lossless electromagnetic environments feature sharp peaks in the spectral response, e.g. a one-dimensional Fabry-P\'erot cavity will be described by a set of harmonic oscillators. In order to allow a reliable numerical description of Fabry-P\'erot cavities, we calculated the Green tensor via
\begin{align*}
g_0(\omega) = \frac{2}{c^2 V} \mathcal{F}_t e^{-\eta t} \sum_{\textbf{k}} \omega_\textbf{k}^{-1} \sin(\omega_\textbf{k} t)~.
\end{align*}
An apparent but critical feature emerges when obtaining $g(\omega)$. The frequency resolution has to be sufficient to properly resolve $g_0$ and $\chi$, both can become sharp for small losses. It is therefore important to converge the spectral representation of $g(\omega)$, for instance by artificially extending the time put into the construction of $g(\omega)$.
A good indication of under-convergence for the Fabry-P\'erot example is when the imaginary part of $g_0(\omega)$ (Lorentzian shape) changes sign around a resonance.
In this work, we increased the frequency resolution, compared to the natural resolution given by the propagation-time, by a factor of 2000 for all reactivity calculations. The Green tensor $G(\omega)$ and $G(t)$ in figure~\ref{fig:reactivity_greensfunction_largeN} have been obtained with a 150 times higher resolution as they are visually identical from this point on and have only been used for visualization. Note, that this resolution is not sufficient to obtain an accurate description of the rather small and thus highly sensitive proton-tunneling process.
Linear response calculations have been performed with a 10 times higher resolution as they feature already from the start a higher resolution. All GPAW calculations used a 1000 times higher resolution. 
The time-stepping remained unchanged such that the $\mathcal{F}_t^{-1} \textbf{G}(\omega)$ and the TDDFT propagation are consistent on the same time-interval. The additional computational cost can become noticeable if the resolution is pushed to extreme values as the FFT becomes then time-consuming. For most applications the cost will remain small or even negligible  as FFTs are well optimized and $\textbf{G}$ is constructed only once before the actual TDDFT calculation. 

The subsequent real-time object $\mathcal{F}_t^{-1} i\omega \textbf{G}(\omega)$ can be contaminated with overtones close to $t=0^+$, the explicit variant $\mathcal{F}_t^{-1} i\omega \textbf{G}(\omega) = -\partial_t \mathcal{F}_t^{-1} \textbf{G}(\omega)$ is preferred. Attention should be devoted to the different possible definitions of the Fourier transformation and their discrete counterparts.
The macroscopic $\chi_E $ is represented either from first-principles or via a Drude-Lorentz model with parameters as indicated in the captions. 
The presented cross sections in figure \ref{fig:responseAlensemble_resonant} and \ref{fig:responseAlensemble_detuned} feature artificial sign-flips towards lower and higher energies (seen from the resonances) as consequence of the finite time-interval. The artificial features were ignored here as they appear at small cross section value and do not affect the conclusions. Stronger damping or longer propagation would remove them if deemed necessary. 

Figure~\ref{fig:responseAlensemble_resonant} and \ref{fig:responseAlensemble_detuned} have been calculated on a grid with 301 grid-points, $0.1~a_0$ spacing and 4th-order finite-difference. The linear-response spectrum has been obtained by perturbing the electronic system with a Lorentzian-shaped delta-kick $v(x,t)=-10^{-4} \frac{1}{\pi} \frac{10^{-2}}{(t-1)^2+10^{-4}} x $ (in a.u.) and time-propagation for 800001 time-steps with $\Delta t=0.01$ a.u. and 4th-order Runge-Kutta.

Figure~\ref{fig:responsesoidum} was obtained in multiple steps as described in the main text. The polarizability of a single ensemble sodium dimer was calculated by real-time propagation via $\alpha_{zz}(\omega)=R_z(\omega)/(2\pi K)$ ($K=10^{-5}$ being the kick-strength) using the LCAO routine of GPAW \cite{enkovaara2010electronic,kuisma2015localized} including the radiation-reaction potential \cite{schafer2021shortcut} with a cross-sectional area $A=10^2$ \AA$^2$. The subsequently obtained spectra using the embedding radiation-reaction calculations used the default real-time to spectrum tool of GPAW with a Lorentzian broadening of $0.05~eV$ ($0.08~eV$ for $8Na_2$ with 20~\AA~ distance). All calculations used a spacing of 0.3~\AA, a box of size 24+chain-length$\times 24 \times 24$~\AA$^3$, the atomic p-valence dz basis and the LDA functional. The sodium dimers have a fixed bonddistance of $1.104$ \AA.

All reactivity calculations have been performed on a grid with 301 grid-points, $0.04~a_0$ spacing and 4th-order finite-difference. The particle-mass was adjusted to the proton-mass. The real-time dynamics involved the deformation of the potential energy surface as specified in the main text with 100001 time-steps with $\Delta t=0.5$ a.u. and 4th-order Runge-Kutta.

\bibliography{sponemission} 

\end{document}